\documentclass[aps,prl,twocolumn,superscriptaddress]{revtex4-1}

\usepackage{graphicx}
\usepackage{subfigure}
\usepackage{dcolumn}
\usepackage{bm}
\usepackage{amsmath}
\usepackage{color}
\usepackage{verbatim}
\usepackage{natbib}
\usepackage{verbatim}
\usepackage[colorlinks,
            linkcolor=blue,
            anchorcolor=blue,
            citecolor=blue,
urlcolor=blue
            ]{hyperref}
\usepackage{color}

\begin{document}


\title{Approaching quantum-limited metrology with imperfect detectors by using weak-value amplification}


\author{Liang Xu}
\affiliation{National Laboratory of Solid State Microstructures, Key Laboratory of Intelligent Optical Sensing and Manipulation, College of Engineering and Applied Sciences, and Collaborative Innovation Center of Advanced Microstructures, Nanjing University, Nanjing 210093, China}
\author{Zexuan Liu}
\affiliation{National Laboratory of Solid State Microstructures, Key Laboratory of Intelligent Optical Sensing and Manipulation, College of Engineering and Applied Sciences, and Collaborative Innovation Center of Advanced Microstructures, Nanjing University, Nanjing 210093, China}
\author{Animesh Datta}
\affiliation{Department of Physics, University of Warwick, Coventry CV4 7AL, United Kingdom}
\author{George C. Knee}
\affiliation{Department of Physics, University of Warwick, Coventry CV4 7AL, United Kingdom}
\author{Jeff S. Lundeen}
\affiliation{Max Planck Centre for Extreme and Quantum Photonics, Department of Physics, University of Ottawa, 25 Templeton Street, Ottawa, Ontario K1N 6N5, Canada}
\author{Yan-qing Lu}
\email{yqlu@nju.edu.cn}
\affiliation{National Laboratory of Solid State Microstructures, Key Laboratory of Intelligent Optical Sensing and Manipulation, College of Engineering and Applied Sciences, and Collaborative Innovation Center of Advanced Microstructures, Nanjing University, Nanjing 210093, China}
\author{Lijian Zhang}
\email{lijian.zhang@nju.edu.cn}
\affiliation{National Laboratory of Solid State Microstructures, Key Laboratory of Intelligent Optical Sensing and Manipulation, College of Engineering and Applied Sciences, and Collaborative Innovation Center of Advanced Microstructures, Nanjing University, Nanjing 210093, China}

\date{\today}

\begin{abstract}
Weak value amplification (WVA) is a metrological protocol that amplifies ultra-small physical effects. However, the amplified outcomes necessarily occur with highly suppressed probabilities,  leading to the extensive debate on whether the overall measurement precision is improved in comparison to that of conventional measurement (CM). Here, we experimentally demonstrate the unambiguous advantages of WVA that overcome practical limitations including noise and saturation of photo-detection and maintain a shot-noise-scaling precision for a large range of input light intensity well beyond the dynamic range of the photodetector. The precision achieved by WVA is six times higher than that of CM in our setup. Our results clear the way for the widespread use of WVA in applications involving the measurement of small signals including precision metrology and commercial sensors.
\end{abstract}

\pacs{}

\maketitle


\textit{Introduction}. The precision of optical metrology and sensing is ultimately determined by the quantum fluctuations of light. Quantum-optical states ($e.g.$, N00N states and squeezed states) can improve the precision of parameter estimation from the shot-noise limit (SNL) \cite{braunstein1992quantum} to the Heisenberg limit (HL) \cite{walther2004broglie, xiao1987precision}. However, such quantum states are very vulnerable to experimental imperfections and are difficult to prepare, especially for large photon numbers \cite{giovannetti2011advances, demkowicz2012elusive}. Instead, a typical approach to enhance precision is to increase the average photon number $\bar{n}$ of the coherent state. In principal, this scheme can attain a precision at SNL, which scales as $1/\sqrt{\bar{n}}$. In practice, this scaling is a challenge due to the ubiquitous noise of detectors \cite{janesick2001scientific}. In particular, the saturation of detectors sets a tight limit on the intensity of the detected light, beyond which the enhancement in the measurement precision by increasing the light intensity is reduced or even eliminated.

Weak value amplification (WVA), deployed to amplify miniscule physical effects through post-selection \cite{Hosten:2008ih, Dixon:2009eu,Strubi:2013hb,MaganaLoaiza:2014kf,Xu:2013fu}, has the potential for enhancing measurement sensitivity and overcoming certain environmental disturbances \cite{Pang:2015ja, Pang:2016hx, Kedem:2012gl,Starling:2009bb,Pang:2015jn,Nishizawa:2012gk,Jordan:2014jv}. Yet to date most of works demonstrating the metrological advantages of WVA are attained under theoretical assumptions and experimental conditions different from those of conventional measurement (CM) \cite{torres2016weak,Combes:2014hd,Vaidman:2017bi,Lee:2014er,Ferrie:2014gf,Viza:2015fj,Brunner:2010dta,Knee:2014dd,Dressel:2014ks,Knee:2014jt,Knee:2018jkb}. Identifying the unambiguous advantage of WVA is still under exploration. With ideal setups, WVA can achieve as good precision as CM \cite{Zhang:2015ko,Tanaka:2013dt, Alves:2015gk}. Crucially, this implies the small number of post-selected photons contain almost all of the metrological information. As a result, WVA potentially provides an approach to ensure that the detector operates under the saturation threshold even for a large number of input photons, thereby preserving the shot-noise-scaling precision and outperforming CM \cite{Harris:2017dg}.

\begin{figure*}
\centering
\includegraphics[width=0.95\textwidth]{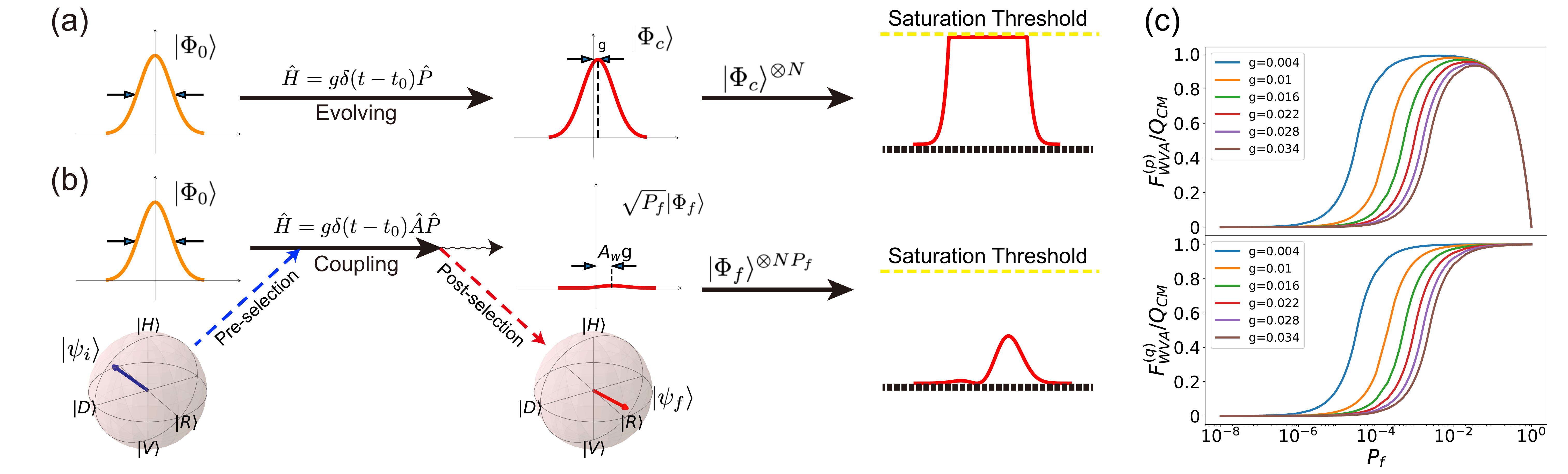}
\caption{Comparison between the conventional measurement and the weak value amplification. (a), Schematic for the conventional measurement (CM) and (b), weak value amplification (WVA) of parameter $g$ with a Gaussian meter state in the position degree of freedom of photons. With a large number of input photons, detector saturation causes distortion in the measurement outcome in CM, while WVA avoids the saturation due to the reduced photon number with post-selection. The upper (lower) figure in (c) plots the ratio of the maximal classical Fisher information $F^{(p)}_{\text{WVA}}$ ($F^{(q)}_{\text{WVA}}$) in WVA with completely imaginary (real) weak value to the quantum Fisher information $Q_{\text{CM}}$ of CM
with ideal detection as a function of the successful post-selection probability $P_f$ for different values of $g$. The width of the Gaussian meter state is $2\sigma=1$.
\label{Schematic}}
\end{figure*}

In this work, we demonstrate the capability of the WVA scheme to overcome the precision limit set by the saturation of the detectors. As an example, we experimentally measure a small transverse displacement of an optical beam, which plays an important role in many applications \cite{rugar1990atomic, howell2006handbook}. The results confirm that WVA outperforms the CM in terms of precision in the presence of detector noise and saturation. Moreover, the optimal precision of WVA can be attained with a widely tunable probability of post-selection, which allows the precision to maintain the shot-noise scaling ($i.e.$, $1.19$ times SNL) for a much larger number of input photons, and extends the dynamic range of the measurement system by two orders of magnitude. Our analysis is also applicable to the measurement of other physical parameters \cite{strubi2013measuring, magana2014amplification, xu2013phase} with different kinds of photodetectors.

\textit{Theoretical Framework.} Fig. \ref{Schematic} describes the measurement of the displacement $g$ with a standard Gaussian meter state (MS) $|\Phi_0\rangle = \int dq\ 1/(2\pi\sigma^2)^{1/4}\exp{[-q^2/(4\sigma^2)]}|q\rangle = \int dp\ (2\sigma^2/\pi)^{1/4}\exp{(-\sigma^2p^2)}|p\rangle$, where $|q\rangle$ and $|p\rangle$ are the eigenstates of the position operator $\hat{Q}$ and the momentum operator $\hat{P}$, respectively. In CM, this meter state is evolved under the Hamiltonian $\hat{H} = g\delta(t-t_0)\hat{P}$, which leads to the final state $|\Phi_c\rangle = \int dq\ 1/(2\pi\sigma^2)^{1/4}\exp{[-(q-g)^2/(4\sigma^2)]}|q\rangle$ with a displacement $g$ in $q$. In contrast, WVA is regarded as an ancilla-assisted metrological scheme. A two-level quantum system (QS) with pre-selected state $|\psi_i\rangle = \cos(\theta_i/2) |0\rangle + \sin(\theta_i/2)e^{i\phi_i} |1\rangle$ is coupled to the meter state by the Hamiltonian $\hat{H} = g\delta(t-t_0)\hat{A}\hat{P}$ and then projected onto the post-selected state $|\psi_f\rangle = \cos(\theta_f/2) |0\rangle + \sin(\theta_f/2)e^{i\phi_f} |1\rangle$, resulting in the final MS $|\Phi_f\rangle$ with the success probability $P_f$, where $\hat{A}$ is an observable of QS. In the weak interaction regime ($g\ll \sigma$), the average shift of $|\Phi_f\rangle$ in $q$ or $p$ space are respectively approximated as $g\text{Re}(A_w)$ and $g\text{Im}(A_w)/(2\sigma^2)$, where $A_w$ is the `weak value' of the observable $\hat{A}$, given by
$A_w=\langle \psi_f|\hat{A}|\psi_i\rangle/\langle \psi_f|\psi_i\rangle$. When the denominator $\langle\psi_f|\psi_i\rangle$ becomes small, $A_w$ can become large giving rise to the amplification effect.

\begin{figure*}[ht]
\centering
\includegraphics[width = 0.95\textwidth]{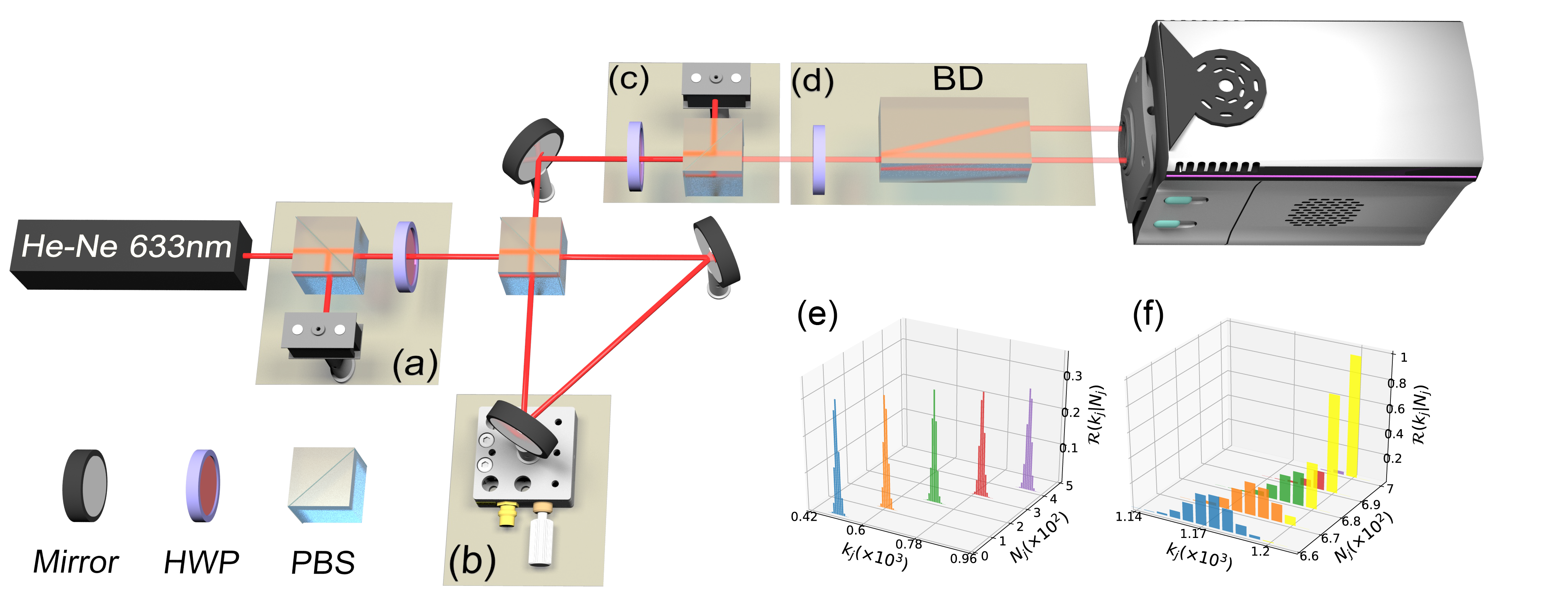}
\caption{Experimental setup and detector calibration. Module (a) and (c) perform the pre- and post-selection of the polarization of photons. The displacement of the mirror in module (b) couples the polarization and the spatial degree of freedom. A HWP and a polarizing beam displacer (BD) in module (d) create two copies of the output beam in order to cancel out the adverse effects of beam jitter and turbulence. Plots (e) and (f) give examples of the measured response matrix of the CCD, \textit{i.e.}, the probability distribution $\mathcal{R}(k_j|N_j)$ of pixel readout $k_j$ when $N_j$ photoelectrons are generated, far from and close to saturation, respectively.
\label{WVA_setup}}
\end{figure*}

To acquire the information about $g$, we perform measurement on the final states $|\Phi_c\rangle$ of CM and $|\Phi_f\rangle$ of WVA, respectively. According to the Cram\'er-Rao bound (CRB), the best precision of estimating $g$ from $\nu$ times of repetitive measurement is given by $\delta^2 g \geq 1/(\nu F)$, where $\delta^2 g$ is the variance of the estimator of $g$ and $F$ is the Fisher information (FI) \cite{Cramer}. The maximum FI, known as the quantum Fisher information (QFI), can be achieved with the optimal measurement on the state. For CM, a measurement in the position $q$ on the $|\Phi_c\rangle$ is optimal such that the FI $F_{\text{CM}}$ of the measured distributions equal the QFI $Q_{\text{CM}} = 1/\sigma^2$. The QFI of WVA $Q_{\text{WVA}}$ depends on the $|\psi_i\rangle$ and $|\psi_f\rangle$ but the maximal $Q_{\text{WVA}} = Q_{\text{CM}}$. In addition, the measurement on $|\Phi_f\rangle$ in the $q$ ($p$) space proves to be optimal if the weak value is completely real (imaginary) such that the FI $F^{(q)}_{\text{WVA}} = Q_{\text{WVA}}$ ($F^{(p)}_{\text{WVA}} = Q_{\text{WVA}}$) \cite{WVA_supplementary}. Therefore, both CM and WVA are optimized, leading to $F_{\text{CM}} = F_{\text{WVA}}$. However, a large number of input photons is more likely to saturate the detectors in CM than in WVA, as illustrated in Fig. \ref{Schematic}, which causes distortion of the measurement on $|\Phi_c\rangle$ and diminishes $F_{\text{CM}}$. This provides a potential advantage of WVA over CM.

We denote the WVA with completely real and completely imaginary weak values as RWVA and IWVA, respectively. Here, our aim is to acquire as high the precision as possible while detecting a limited number of photons. We maximize the FI $F_{\text{WVA}}^{(q)}$ ($F_{\text{WVA}}^{(p)}$) normalized to $Q_{\text{CM}}$ over a range of $P_f$ for different $g$ \cite{WVA_supplementary}, which is shown in Fig. \ref{Schematic} (c). The RWVA shows obvious advantages in that the increase of $P_f$ always promises an enhanced precision. For example, with parameters $g=0.01$ and $\sigma=0.5$ in the RWVA scheme, $F_{\text{WVA}}^{(q)}$ can attain over $99\%$ of $Q_{\text{CM}}$ over a large range ($1\%$ to $100\%$) of the input photons being detected. This incredible property provides us with great flexibility for the choice of $P_f$. Consequently, WVA can operate in an intensity range well above the noise floor but under the saturation level of detectors, and simultaneously, maintain the metrological information. By contrast, there is a peak value of $F_{\text{WVA}}^{(p)}/Q_{\text{CM}}$ with the change of $P_f$ in IWVA. Indeed, Ref. \cite{Li:2017bx} also shows that in IWVA, the optimal choice of $P_f$ (and, hence, the pre- and post-selected states) is sensitive to the parameter $g$. It follows that one must have some prior knowledge of $g$ in order to design a measurement system using IWVA. Consequently, we choose to implement the optimal RWVA scheme and make a comparison to CM in our experiment. 

\textit{Experiment.} The experimental setup is shown in Fig. \ref{WVA_setup}. The polarization and spatial degrees of freedom of the CW laser beam at 633nm with the TEM00 mode (beam width $\sigma = 0.472$mm) are used as the ancillary QS ($|0\rangle\rightarrow |H\rangle, |1\rangle\rightarrow |V\rangle$) and the Gaussian MS, respectively. The input photons pass through a polarizing beam splitter (PBS) and a half-wave plate (HWP) to prepare the pre-selected state. The photons in $|H\rangle$ ($|V\rangle$) state go clockwise (anti-clockwise) in the Sagnac interferometer. A slight displacement of the mirror results in the coupling between the QS and the meter state. After recombination at the output port, a HWP and a PBS performs the post-selection. The meter state is then measured in the position $q$ space by a scientific CCD (Andor, iStar CCD 05577H) with pixel size $13\times 13 \mu m$. If the total average number of photons in the Gaussian beam is $\bar{n}_t$ per exposure, the $j\text{th}$ pixel of the CCD is expected to receive $\bar{n}_j^{\text{WVA}}(g)=P_f\bar{n}_t\int_j dq |\langle q|\Phi_f\rangle|^2$ and $\bar{n}^{\text{CM}}_j(g)=\bar{n}_t\int_j dq|\langle q|\Phi_c\rangle|^2$ photons in the WVA and CM schemes, respectively. Since the beam is in a coherent state, the exact number $N_j$ of the registered photons ($i.e.$, photoelectrons) at the $j\text{th}$ pixel follows a Poisson distribution $P(N_j|\eta\bar{n}_j,g)$, where $\eta=0.125$ is the detection efficiency of the CCD. Additionally, due to various kinds of electrical noise, the response of CCD can be described by a conditional probability distribution $\mathcal{R}(k_j|N_j)$, where $k_j$ is the readout at the $j$th pixel. The readout $k_j$ contains the contributions from the dark noise $K_d$, $N_j$, and the extra classical noise $K_a$. The calibration shows that $K_d$ and $K_a$ follow normal distributions $K_d \sim \mathcal{N}(\mu_d, \sigma^2_d)$ and $K_a \sim \mathcal{N}(0, \sigma^2_a)$, in which $\sigma_a$ grows with $\bar{n}_j$, following $\ln (\sigma_a^2) = a\ln (\bar{n}_j)+b$ with $a=1.19$ and $b=-4.39$. Thus, $\mathcal{R}(k_j|N_j)$ is obtained by the convolution of dark noise and the extra classical noise distributions. Given the saturation threshold $k_s$, the response at the threshold is transformed to $\mathcal{R}(k_s|N_j) = \sum_{k_j\ge k_s} \mathcal{R}(k_j|N_j)$. We present certain $\mathcal{R}(k_j|N_j)$ of our detectors in Fig. \ref{WVA_setup} (e) and (f). The response model of CCD here is similar to that in Ref. \cite{Harris:2017dg} with digitalization and pixel noise.

\begin{figure*}[t]
\centering
\includegraphics[width=1\textwidth]{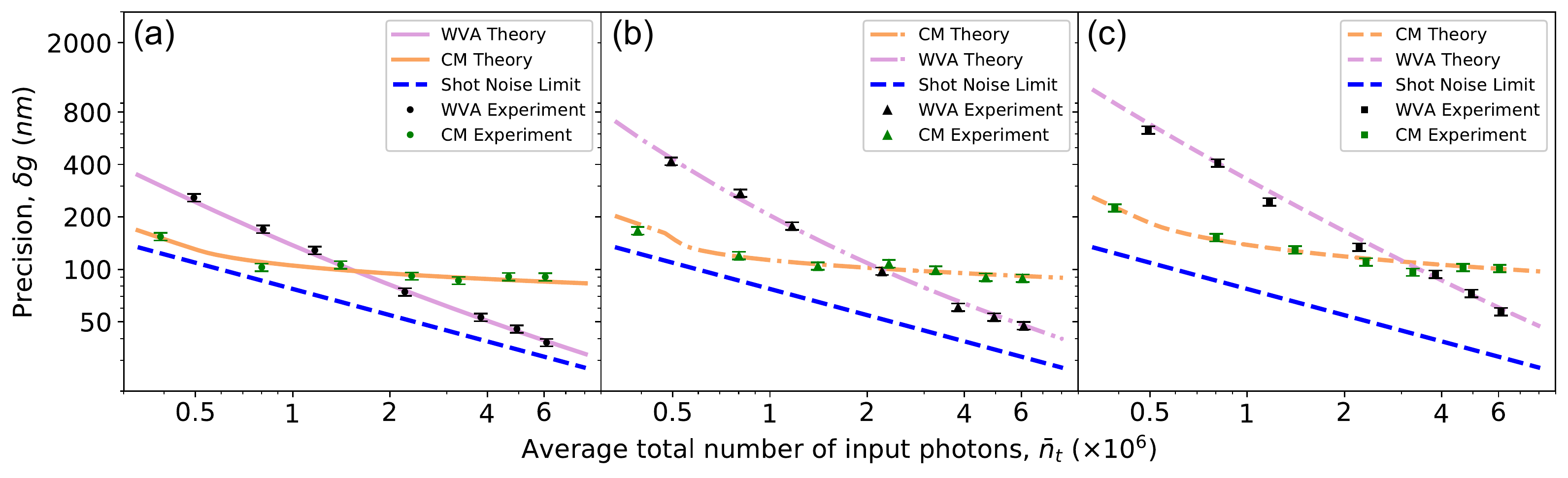}
\caption{Comparison between the precision of CM and that of WVA in the estimation of $g$. (a), Maximum likelihood estimation (MLE). (b), split-detection (SD) estimation. (c), center-of-mass (COM) estimation. The theoretical results of MLE are determined by the Cram\'er-Rao bound while the theoretical lines in SD and COM are derived by the error propagation formula. All error bars refer to $\pm 1$ s.d. and are calculated from the fourth moment of the estimated parameter $g$. The 'shot noise limit' is determined by $\delta g = \sigma/\sqrt{\nu \eta \bar{n}_t}$. 
\label{Precision_Results}}
\end{figure*}

\begin{figure}[hb]
\centering
\includegraphics[width=0.45\textwidth]{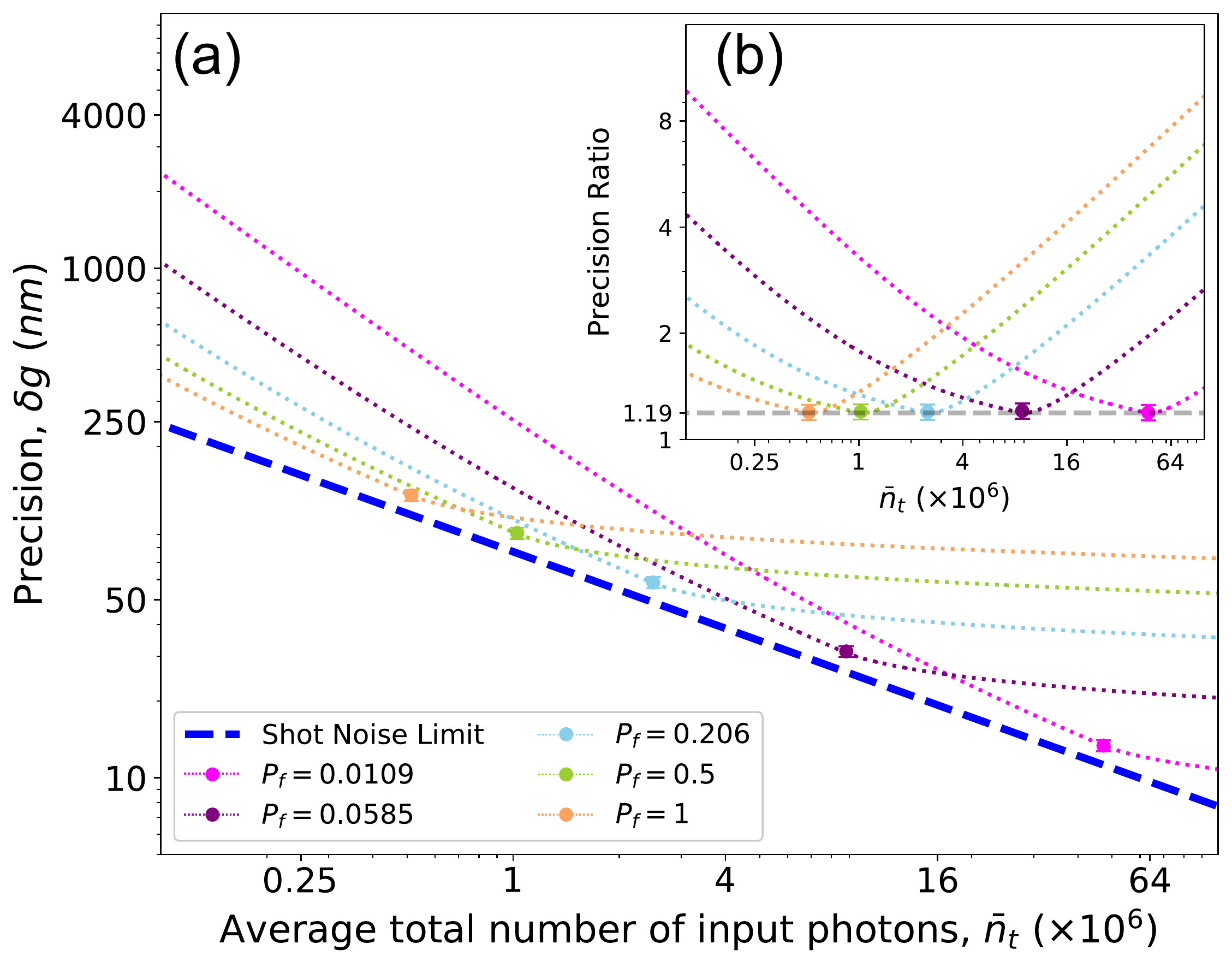}
\caption{Weak value amplification (WVA) with different success probabilities of post-selection $P_f$. (a), shows the precision of WVA with different $P_f$ and average total number of input photons $\bar{n}_t$. The error bars refer to $\pm 1$ s.d.. In (b), we plot the ratio of the precision in WVA to the shot-noise limit (SNL). In our experiment, we implement the optimal precision WVA scheme which has the minimum ratio $1.19$ for each specified $\bar{n}_t$. The theoretical results (dotted lines) are determined by the Cram\'er-Rao Bound and the experimental results (points) are obtained using Maximum likelihood estimation. 
\label{SNS_Results}}
\end{figure}

Taking all of the mentioned factors into consideration, the conditional probability distribution of $k_j$ that depends on $g$ is given by,
\begin{equation}\label{p.k.g}
P(k_j|g) = \sum_{N_j}\mathcal{R}(k_j|N_j)P(N_j|\eta\bar{n}_j,g).
\end{equation}
Subsequently, the classical Fisher information for $g$ of the whole CCD array can be calculated from $P(k_j|g)$ and simplified as
\begin{equation}\label{FI_t}
F(g)=\sum_j \frac{\eta}{\bar{n}_j}\left(\frac{d\bar{n}_j}{dg}\right)^2 \Gamma(\mathcal{R}, \bar{n}_j),
\end{equation}
in which the coefficient $\Gamma(\mathcal{R}, \bar{n}_j)$ can be treated as the signal-to-noise ratio at the $j$th pixel, taking into account the fundamental quantum fluctuations of light and probabilistic response of the CCD. When the $\bar{n}_j$ in Eq. \eqref{FI_t} is replaced by $\bar{n}_j^{\text{CM}}$ or $\bar{n}_j^{\text{WVA}}$, we can obtain the FI of CM or WVA, respectively. Eq. \eqref{FI_t} also implies that the response of the CCD $\mathcal{R}(k_j|N_j)$ plays a vital role in acquiring the Fisher information $F(g)$. For example, when the pixel approaches saturation, $\Gamma(\mathcal{R}, \bar{n}_j)$ tends to zero, leading to a considerable reduction in FI \cite{WVA_supplementary}.

Three methods are employed to estimate the parameter $g$, Maximum Likelihood estimation (MLE), split-detection (SD) estimation, and center-of-mass (COM) estimation. In MLE, the parameter $g$ is estimated by maximizing the likelihood function 
\begin{equation}
\mathcal{L}(g)=\prod_{l=1}^{\nu}\prod_{j=1}^{\tau}\left[\sum_{N_{lj}}\mathcal{R}_s(k_{lj}|N_{lj})P(N_{lj}|\eta\bar{n}_{lj},g)\right],
\end{equation}
where $\nu=300$ and $\tau=330$ are the total number of frames and pixels in one estimate, respectively. $\nu$ frames are randomly selected from a set of 6000 frames using a bootstrap method, which repeats 200 times to obtain $\delta g_{\text{MLE}}$. Since MLE is known to be able to saturate the Cram\'er-Rao bound asymptotically, the theoretical precision of WVA is given by $\delta^2 g_{\text{MLE}} = 1/[\nu F(g)]$. Given the center coordinate of the initial MS $X_0$, the COM estimator is formulated as $\hat{g}_{\text{COM}}=\sum_{j=1}^{\tau}w_jx_j-X_0$, where $x_j$ and $w_j = (k_j-\mu_d)/\sum_{j=1}^{\tau} (k_j-\mu_d)$ are the coordinate and the normalized weight of the $j$th pixel, respectively. Thus, $\delta g$ in COM estimation can be inferred from $\delta^2 g_{\text{COM}}=\sum_{j=1}^{\tau}\delta^2w_jx^2_j/\nu$. In SD estimation, a row of pixels is divided into two sections and taken as a split detector by summing up the pixel readouts and subtracting the dark counts of each section. The normalized results of the entire left and right sections are $W_l$ and $W_r$. The SD estimator is $\hat{g}_{\text{SD}}=(W_l-W_r)/\xi$ and the variance of $g$ is $\delta^2 g_{\text{SD}}=(\delta^2 W_l+\delta^2 W_r)/(\nu \xi^2)$. The coefficient $\xi$ equals $\sqrt{2/\pi}/\sigma$ before the detector saturation and decreases when saturation occurs \cite{WVA_supplementary}.

We first compare the precision of CM and WVA with the identical MS for a range of the average total number of input photons $\bar{n}_t$. In WVA, we set $\theta_i = -\theta_f = 76^\circ$, $P_f = 0.0585$ and $A_w = 4.13$. From Fig. \ref{Precision_Results}, the obtained precision of all the three estimation methods follow similarly varying trends. CM outperforms WVA when $\bar{n}_t$ is small because CM collects all the input photons, which helps diminish the impact of dark noise. However, as $\bar{n}_t$ gets large, saturation begins to occur in the CM scheme, thereby restricting further improvements of precision. In contrast, by concentrating the metrological information into many fewer photons, WVA avoids saturation and maintains the increase of  precision for a large $\bar{n}_t$. We also give an intuitive illustration in which the FIs distributed in each pixel of CCD with the increase of total input photons are compared between CM and WVA \cite{WVA_supplementary}. For both CM and WVA, the precision of three methods - MLE, SD, COM, decreases in this order, which conforms to our intuition a more accurate model that takes into account more characteristics of the experimental apparatus, tends to extract more information from a particular probability distribution.

Furthermore, we have also compared the precision of WVA with $P_f = 0.0109, 0.0585, 0.206, 0.5, 1$ by setting $\theta_i = -\theta_f = 84^\circ,76^\circ,63^\circ,45^\circ,0^\circ$ and $\phi_i = \phi_f = 0$, respectively. The MLE results are shown in Fig. \ref{SNS_Results}. We find that for a specified average number of input photons $\bar{n}_t$, there exists the optimal choice of $P_f$ to achieve the best precision due to the trade-off between resisting the various types of noise and avoiding saturation in photodetectors. In our experiment, we adapt the $P_f$ and obtain the optimal precision for each $\bar{n}_t$ with MLE, which is $1.19$ times the SNL. Besides, SD and COM estimators give the optimal precision $1.29$ and $1.67$ times SNL.

\textit{Discussion.} Although keeping a small post-selection probability $P_f$ avoids the detector saturation to a large extent, this may not always be the optimal strategy in the presence of detector noise. The true power of WVA is to adjust $P_f$ over a large range while maintaining the metrological information, which allows to minimize the overall detector imperfections and maximize the precision. This is also the reason for the advantage of RWVA over IWVA since the former provides a greater flexibility for the choice of $P_f$. 

In summary, by preserving all the metrological information with a tunable fraction of post-selected photons, WVA promises to be a key technique for extending the dynamic range of a measurement system. More generally, we have shown that it is possible to channel metrological information into particular aspects of a sensor output signal, thereby allowing one to match them to the detector aspects that have the best specifications ($e.g.$, resolution, noise in a certain intensity range, bit depth, $etc.$). For now, by resolving the controversy around WVA, we have opened a path to its application in a wide variety of commercial, industrial, and scientific sensors and instruments.

\begin{acknowledgments}
The authors thank B.J. Smith and L.L S\'anchez-Soto for assistance and fruitful discussions. This work was supported by the National Key Research and Development Program of China (Grant Nos. 2017YFA0303703 and 2018YFA030602) and the National Natural Science Foundation of China (Grant Nos. 91836303, 61975077, 61490711 and 11690032). A. D. is supported by the UK EPSRC (EP/K04057X/2), the UK National Quantum Technologies Programme (EP/M01326X/1, EP/M013243/1) and the University of Warwick Global Partnership Fund. 
\end{acknowledgments}

\end{document}